\documentclass[9pt,twocolumn,twoside]{pnas-new}

\usepackage{ifthen}
\newcommand\x{0}
\ifthenelse{{\x=1}}{\newcommand\colornewtext{red}}{\newcommand\colornewtext{black}}
\newcommand\colorremovedtext{black}
\definecolor{BlueSigSt}{rgb}{0.0510,0.4235,0.6627}
\ifthenelse{{\x=1}}{\newcommand\colornewtextSignificanceStatement{red}}{\newcommand\colornewtextSignificanceStatement{BlueSigSt}}

\templatetype{pnasresearcharticle} 

\title{Dirac cones and chiral \ifthenelse{{\x=1}}{\textcolor{\colorremovedtext}{\st{waves}}}{}\textcolor{\colornewtext}{selection of elastic waves} in a soft \ifthenelse{{\x=1}}{\textcolor{\colorremovedtext}{\st{material}}}{}\textcolor{\colornewtext}{strip
}}

\author[a,1]{Maxime Lanoy}
\author[a]{Fabrice Lemoult} 
\author[b]{Antonin Eddi}
\author[a]{Claire Prada}

\affil[a]{Institut Langevin, ESPCI Paris, PSL University, CNRS, 75005 Paris, France}
\affil[b]{PMMH, CNRS, ESPCI Paris, Universit\'e PSL, Sorbonne Universit\'e, Universit\'e de Paris, F-75005, Paris, France}

\leadauthor{Lanoy} 

\significancestatement{\textcolor{\colornewtextSignificanceStatement}{Thanks to particle tracking methods, we monitor the propagation of in-plane elastic waves in an incompressible thin strip and observe, for the first time, a Dirac cone in a soft material.}
\ifthenelse{{\x=1}}{\textcolor{BlueSigSt}{\st{We present the first observation of a Dirac cone in a soft material.}}}{} 
\ifthenelse{{\x=1}}{\textcolor{BlueSigSt}{\st{Exploiting the exceptional properties of this cone, we report}}}{} \textcolor{\colornewtextSignificanceStatement}{Additional remarkable wave features such as} negative phase velocities, \textcolor{\colornewtextSignificanceStatement}{pseudo} zero group velocity and \ifthenelse{{\x=1}}{\textcolor{BlueSigSt}{\st{unidirectional chiral waves}}}{}\textcolor{\colornewtextSignificanceStatement}{one-way chiral selection are highlighted}. Our findings are universal\ifthenelse{{\x=1}}{\textcolor{BlueSigSt}{\st{ and give access to a new paradigm}}}{}: any thin strip made of any soft elastomer will display the same behavior.
Dirac cones have inspired many developments in the condensed matter field over the last decade. Our findings enable the search for analogues in the realm of soft matter, leading to a wide range of potential applications. Additionally, they \ifthenelse{{\x=1}}{\textcolor{BlueSigSt}{\st{may have a practical impact}}}{}\textcolor{\colornewtextSignificanceStatement}{are of practical interest} for biologists since soft strips are ubiquitous among human tissues and organs.}

\authorcontributions{All authors participated in the conception of the project, the setting up of the experiment, the data processing and in writing down the article. F.L. and M.L. ran the numerical simulations. C.P. provided the theoretical description. A.E. provided the rheological measurements.}
\correspondingauthor{\textsuperscript{1}To whom correspondence should be addressed. E-mail: maxime.lanoy@gmail.com}


\begin{abstract}
We study the propagation of in-plane elastic waves in a soft thin strip;
a specific geometrical and mechanical hybrid framework which we expect to exhibit Dirac-like cone. We separate the low frequencies guided modes (typically 100~Hz for a \ifthenelse{{\x=1}}{\textcolor{\colorremovedtext}{\st{centimetric}}}{}\textcolor{\colornewtext}{centimetre wide} strip) and obtain experimentally the full dispersion diagram.\ifthenelse{{\x=1}}{\textcolor{\colorremovedtext}{\st{Dirac-like cones are evidenced and go along with remarkable wave features such as negative phase velocities or zero group velocity (ZGV).}}}{}
\textcolor{\colornewtext}{Dirac cones are evidenced together with other remarkable wave phenomena such as negative wave velocity or pseudo-zero group velocity (ZGV)}. Our measurements are convincingly supported by a model (and numerical simulation) for both Neumann and Dirichlet boundary conditions. Finally, we perform \textcolor{\colornewtext}{one-way chiral selection}\ifthenelse{{\x=1}}{\textcolor{\colorremovedtext}{\st{selective generation of one-way chiral modes}}}{} by carefully setting the source position and polarization. Therefore, we show that soft materials support atypical wave-based phenomena, which is all the more interesting as they make most of the biological tissues.
\end{abstract}

\dates{This manuscript was compiled on \today}
\doi{\url{www.pnas.org/cgi/doi/10.1073/pnas.XXXXXXXXXX}}

\begin{document}

\maketitle
\thispagestyle{firststyle}
\ifthenelse{\boolean{shortarticle}}{\ifthenelse{\boolean{singlecolumn}}{\abscontentformatted}{\abscontent}}{}

\dropcap{G}raphene has probably become the most studied material in the last decades. It displays unique electronic properties resulting from the existence of the so-called Dirac cones~\cite{neto2009electronic}. At these degeneracy points, the motion of electrons is described in quantum mechanics by the Dirac equation: the dispersion relation becomes linear and electrons behave like massless fermions~\cite{ashcroft2010solid}. As a result, interesting transport phenomena such as the Klein tunneling or the {\it Zitterbewegung} effect have been reported~\cite{katsnelson2006chiral}. But Dirac cones are not specific to graphene. They correspond to transition points between different topological phases of matter~\cite{hasan2010colloquium}. This discovery has enabled the understanding of topologically protected transport phenomena, such as the quantum Hall effect~\cite{klitzing1980new}.\\
Dirac cones are the consequence of a specific spatial patterning rather than a purely quantum phenomenon. Inspired by these tremendous findings from condensed matter physics, the wave community thus started to search for classical analogs in photonic crystals~\cite{PhysRevB.80.155103,PhysRevB.82.014301}\ifthenelse{{\x=1}}{\textcolor{\colorremovedtext}{\st{ (or phononic crystals when dealing with acoustic waves~}\cite{sukhovich2008negative}\st{)and more recently in metamaterials~}\cite{yves2018measuring}}}{}. Abnormal transport properties similar to the {\it Zitterbewegung} effect were highlighted~\cite{SepkhanovPRA2007,ZhangPRL2008b}. In recent years, the quest for photonic (and phononic) topological insulators~\cite{ozawa2019topological} has become a leading topic. This specific state of matter results from the opening of a band gap at the Dirac frequency and is praised for its application to robust one-way wave-guiding~\cite{khanikaev2013photonic, wu2015scheme}. Surprisingly, similar degeneracies have been observed for unexpected photonic lattices as the consequence of an accidental adequate combination of parameters~\cite{huang2011dirac}. Such Dirac-like cones have a fundamentally different nature as they occur in the $k\to 0$ limit~\cite{mei2012first} but still offer interesting features: wave-packets propagate with a non-zero group velocity while exhibiting no phase variation, just like in a zero-index material~\cite{liu2012dirac, moitra2013realization}.\\
A similar accidental $k\to 0$ Dirac-like cone can be observed in the dispersion relation of elastic waves propagating in a simple plate. In this context, the cone results from the coincidence of two cut-off frequencies occurring when the Poisson's ratio is exactly of $\nu=1/3$~\cite{Mindlin06, maznev2014dirac, stobbe2017, stobbe2019}. This condition seriously restricts the amount of potential materials to nearly the  Duraluminum or zircalloy. However, a recent investigation emphasized that \ifthenelse{{\x=1}}{\textcolor{\colorremovedtext}{\st{, for in-plane waves propagating in a thin strip, the degeneracy is expected for $\nu=1/2$
, that is to say for soft materials }}}{}\textcolor{\colornewtext}{the in-plane modes of a thin strip are analogous to Lamb waves propagating in a plate of Poisson's ratio $\nu'=\nu/(1+\nu)$~\cite{laurent2020}. The degeneracy should then occur in the case of incompressible materials ($\nu=1/2$)}. This \ifthenelse{{\x=1}}{\textcolor{\colorremovedtext}{\st{supposes}}}{}\textcolor{\colornewtext}{indicates} that the strip configuration is the \ifthenelse{{\x=1}}{\textcolor{\colorremovedtext}{\st{key to observe}}}{}\textcolor{\colornewtext}{perfect candidate for the observation of} Dirac cones\ifthenelse{{\x=1}}{\textcolor{\colorremovedtext}{\st{ and decline the associated transport effects}}}{} in the world of soft matter. \textcolor{\colornewtext}{Due to their nearly-incompressible nature, soft materials indeed present interesting dynamical properties embodied by the propagation of elastic waves: the velocity of the transversely polarized waves is several orders of magnitudes smaller than its longitudinal counterpart. This aspect has been at the center of interesting developments in various contexts from evidencing the role of surface tension in soft solids~\cite{shao2018extracting,Antonin} to model experiments for fracture dynamics~\cite{livne2005universality} or transient elastography~\cite{sandrin2002shear,Bercoff2004IEEE}.}\\
In this article, we study in-plane elastic waves propagating in a soft \textcolor{\colornewtext}{(\textit{i.e.} incompressible and highly deformable)} thin strip and propose an experimental platform to \ifthenelse{{\x=1}}{\textcolor{\colorremovedtext}{\st{map}}}{}\textcolor{\colornewtext}{monitor} the \textcolor{\colornewtext}{propagation of the in-plane displacement} \ifthenelse{{\x=1}}{\textcolor{\colorremovedtext}{\st{field displacements}}}{}\textcolor{\colornewtext}{thanks to a particle tracking algorithm}. We provide full experimental and analytical description of these in-plane waves both for free and rigid edge conditions. We notably extract the low-frequency part of the dispersion diagram for \ifthenelse{{\x=1}}{\textcolor{\colorremovedtext}{\st{both}}}{}the \textcolor{\colornewtext}{two} configurations. \ifthenelse{{\x=1}}{\textcolor{\colorremovedtext}{\st{Not only}}}{}\textcolor{\colornewtext}{W}e clearly evidence the existence of Dirac-like cones for \ifthenelse{{\x=1}}{\textcolor{\colorremovedtext}{\st{such a}}}{}\textcolor{\colornewtext}{this} simple geometry \ifthenelse{{\x=1}}{\textcolor{\colorremovedtext}{\st{, but we also}}}{}\textcolor{\colornewtext}{ and} highlight some other remarkable wave phenomena such as backward modes or zero group velocity (ZGV) modes. Eventually, we\ifthenelse{{\x=1}}{\textcolor{\colorremovedtext}{\st{exploit these features to}}}{} perform \textcolor{\colornewtext}{chiral} selective excitation resulting in \textcolor{\colornewtext}{the propagation of} \ifthenelse{{\x=1}}{\textcolor{\colorremovedtext}{\st{a}}}{}one-way \textcolor{\colornewtext}{state,} \ifthenelse{{\x=1}}{\textcolor{\colorremovedtext}{\st{propagation of chiral modes}}}{} and in the separation of the two contributions of a ZGV wave.\\

\begin{figure}[bt]
\includegraphics[width=8.7cm]{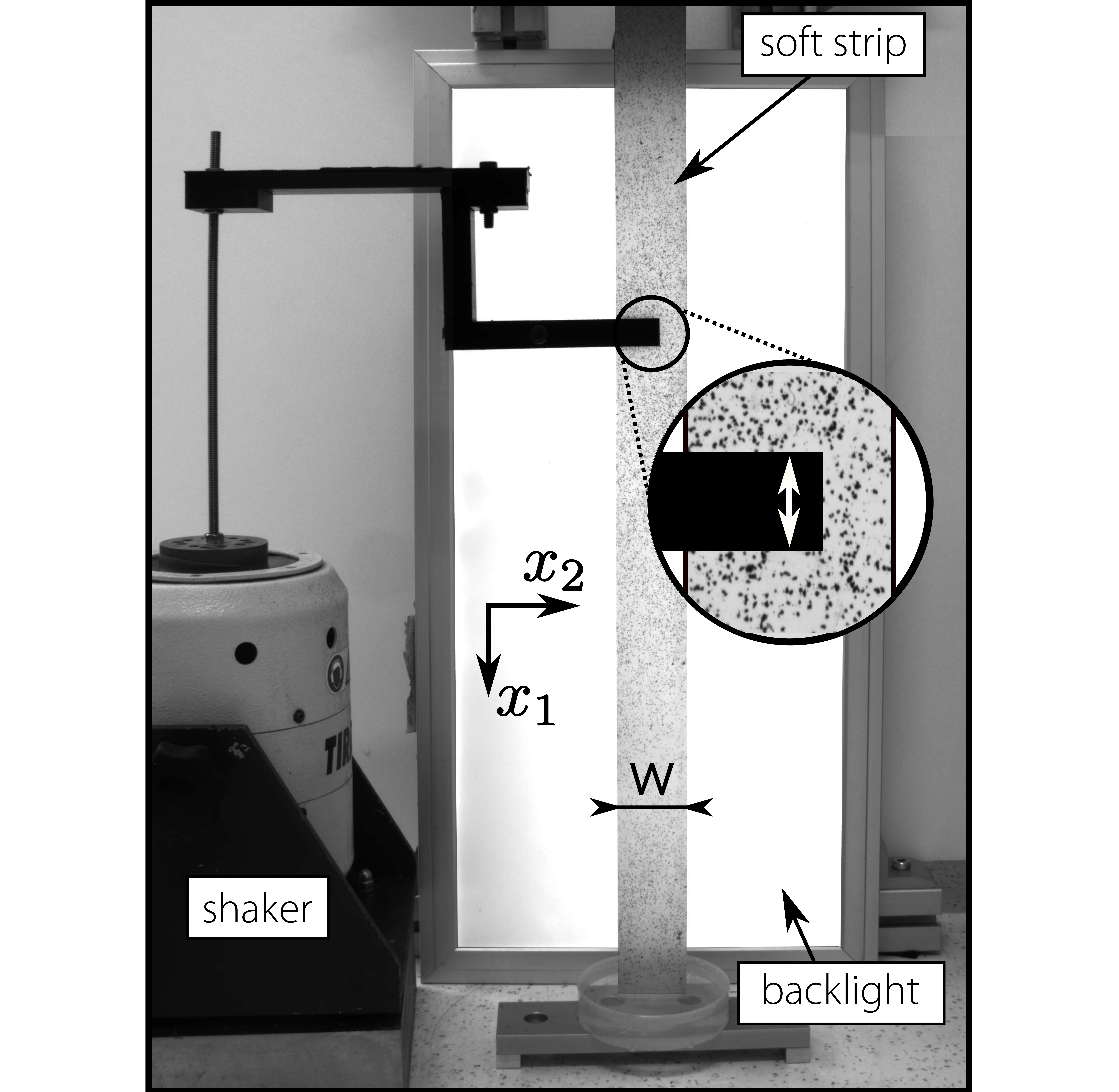}
\caption{\label{fig_1} \textbf{Experimental setup:} a soft elastic strip (of dimensions $L=600$~mm, $w=39$~mm, $d=3$~mm) seeded with dark pigments (for motion tracking purposes) is suspended. A shaker connected to a clamp induces in-plane displacement propagating along the strip.}
\end{figure}

\begin{figure*}[bt]
\vspace{20pt}
    \includegraphics[width=17.8cm]{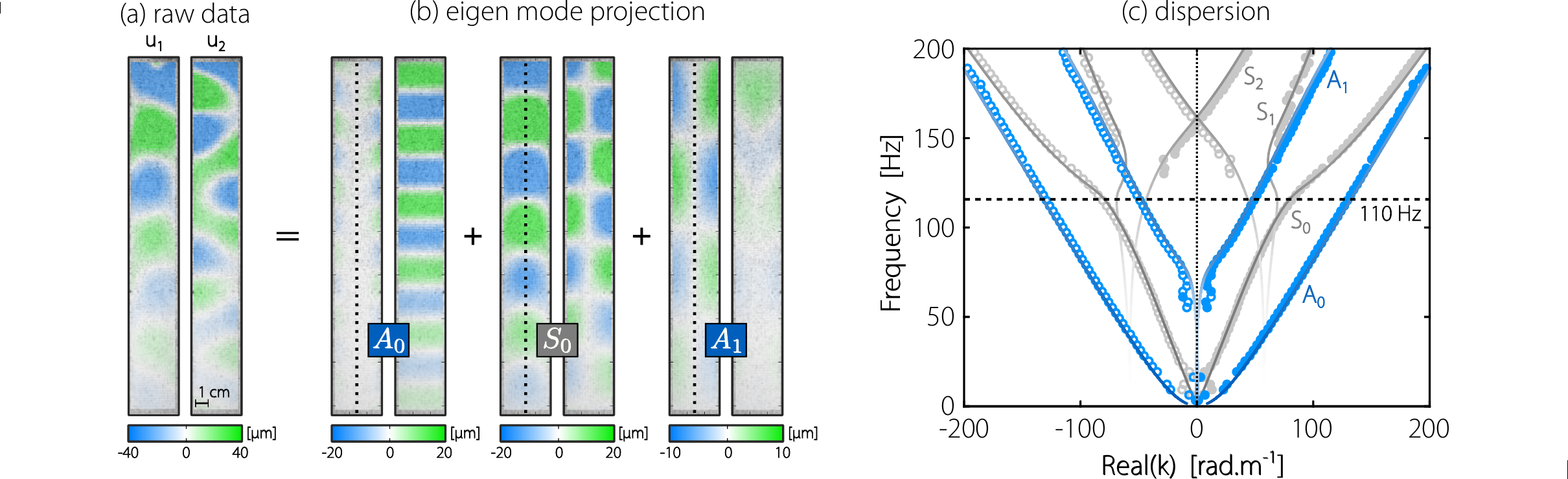}
\caption{\textbf{Free edges field maps and dispersion.} Here $w=39$~mm. (a) Real part of the raw displacements at $110$~Hz and (b) the three corresponding singular vectors (see text). (c) Experimental (symbols) and analytical (solid lines) dispersion curves. Transparency renders the ratio $\text{Im}(k)/\text{Abs}(k)$ \textcolor{\colornewtext}{(see Supplementary Information)}. Filled gray and blue symbols correspond to extracted symmetrical and anti-symmetrical modes. Empty ones are obtained by symmetry.} \label{fig_2}
\end{figure*}

\begin{figure}[t]
\includegraphics[width=8.7cm]{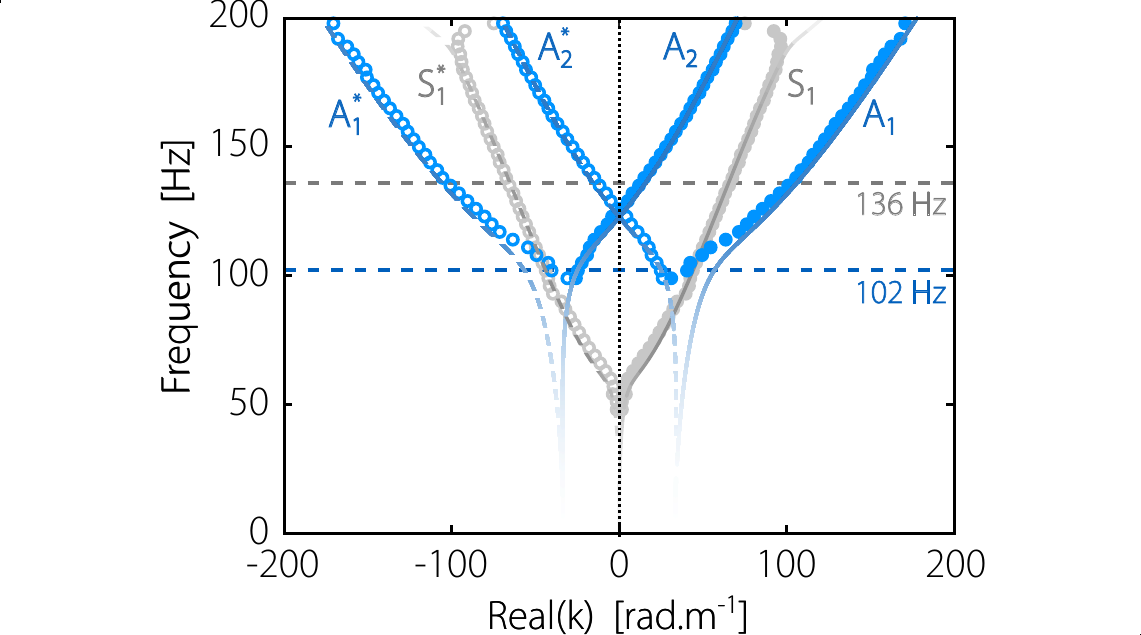}
\caption{\label{fig_3} \textbf{Fixed edges dispersion.} Experimental (symbols) and theoretical (solid lines) dispersion curves for a strip of width $w=50.6$~mm with fixed edges. Symmetrical modes (resp. anti-symmetrical) are labelled in gray (resp. blue). \textcolor{\colornewtext}{Similarly to Fig.~\ref{fig_2}.c, the transparency renders the ratio $\text{Im}(k)/\text{Abs}(k)$ (see Supplementary Information). Filled gray and blue symbols correspond to extracted symmetrical and anti-symmetrical modes. Empty ones are obtained by symmetry.}}
\end{figure}

\begin{figure*}
\vspace{20pt}
    \includegraphics[width=17.8cm]{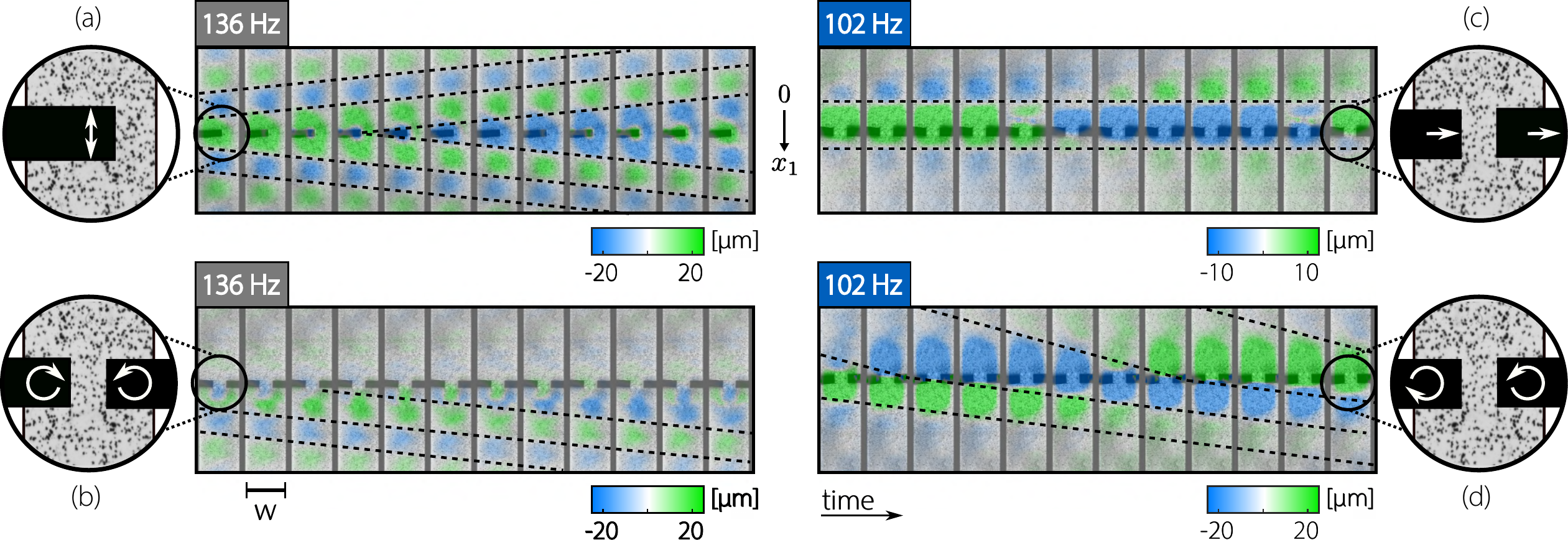}
\caption{\label{fig_4} \textbf{Selective generation.} Chronophotographic sequences (12 snapshots) over a full oscillation cycle. (a) The source is \textcolor{\colornewtext}{placed} at the centre of the strip and shaken vertically at 136~Hz: symmetric diverging waves are observed on both parts. (b) Two sources facing each other are rotated \textcolor{\colornewtext}{in opposite directions at 136~Hz: the wave only travels to the $x_1>0$ region.}
(c) Two sources are shaken horizontally at 102~Hz: a stationary wave associated to an anti-symmetric \textcolor{\colornewtext}{pseudo-}ZGV mode is observed. \textcolor{\colornewtext}{(d) The two sources are rotated at 102~Hz in an anti-symmetrical manner: The propagation is restored and the phase velocity is negative in the on the top region ($x_1<0$). The black dashed lines are visual guides highlighting the zeroes of displacement and the sketches show the source shape and motion. For sake of clarity, one only represents $u_1$ for (a) and (b) and $u_2$ for (c) and (d). See videos S6 to S9 in supporting information for more details.} 
}
\end{figure*}

\ifthenelse{{\x=1}}{\textcolor{\colorremovedtext}{\st{\textbf{Material} \textbf{and} \textbf{methods}. We realize soft thin strips with a silicone elastomer (Smooth-On Ecoflex\textsuperscript{\textregistered} 00-30). The monomer and cross-linking agent are mixed in a 1:1 ratio and left for curing during approximately half a day. Meanwhile, the blend is seeded with dark pigments (smaller than 500~$\mu$m) for tracking purposes. Once cured, the strip dimensions are of $L\times w\times h=600\textrm{ mm}\times39\textrm{ mm}\times3\textrm{ mm}$ and the polymer is found to have a density of $\rho=1010~\textrm{kg.m}^{-3}$. The strip is then suspended with its lower end immersed in a glycerol container to avoid spurious reflections and out-of plane motions (Fig.}~\ref{fig_1}\st{). Indeed, all along the article, we will focus on the in-plane vibrations (\textit{i.e.} displacements components $u_1$ and $u_2$ corresponding respectively to directions $x_1$ and $x_2$) resulting in a guided propagation along the vertical direction~$x_1$. A 3D-printed clamp tightens the strip at a specific location and is designed with conical termination in order to ensure point-like actuation. The clamp is mounted on a shaker (Tira Vib 51120) driven monochromatically by an external signal generator and amplifier (Tira Analog Amplifier BAA 500) with frequencies ranging from $1$ to $200~$Hz. The in-plane displacements are monitored \textit{via} a digital camera (Basler acA4112-20um) positioned approximately 2 meters away from the strip (see video in Supporting Information). Stroboscopic imaging (60 images per period) is performed by slightly detuning the frame rate with respect to the excitation frequency. The video data is then processed with a Digital Image Correlation (DIC) algorithm}~\cite{DIC_Sander}\st{ in order to track the motion of the dark seeds. Complex monochromatic displacement fields are hence mapped on $620\times4112$ pixels. Typical displacement fields ($u_1$, $u_2$) measured when shaking at 110~Hz are reported on Fig.}~\ref{fig_2}\st{(a). This method is sensitive to displacement magnitudes in the micrometer range and enables the probing of the displacement within the whole scanned area in spite of the significant viscous damping expected for such silicone polymers.
}}}{}

\matmethods{
\textcolor{\colornewtext}{\textbf{Sample preparation.} 
The strips are prepared by molding a commercial elastomer (Smooth-On Ecoflex\textsuperscript{\textregistered} 00-30). The monomer and cross-linking agent are mixed in a 1:1 ratio and left for curing for roughly half a day. Once cured, the measured polymer density is of $\rho=1010~\textrm{kg.m}^{-3}$. Rheological measurements are performed on a conventional apparatus (Anton-Paar MCR501) set in a plate-plate configuration. The results are available in Supporting Information.}

\textcolor{\colornewtext}{\textbf{Vibration.} The strips are excited by a shaker (Tira Vib 51120) driven monochromatically with an external signal generator (Keysight 33220A) and amplifier (Tira Analog Amplifier BAA 500) with frequencies ranging from $1$ to $200~$Hz. A point-like excitation is ensured by connecting the shaker to a 3D-printed clamp tightening the strip at a specific location and designed with conical termination. Spurious out of plane vibrations are reduced by immerging the strip's bottom end in glycerol (visible in figure 1).}

\textcolor{\colornewtext}{\textbf{Motion tracking.} During the curing stage, the blend is seeded with "Ivory black" dark pigments (the particles are smaller than 500~$\mu$m) enabling to monitor the motion by Digital Image Correlation (DIC). Video imaging is performed with a wide-sensor camera (Basler acA4112-20um) positioned roughly 2 meters away from the strip (raw videos are available in Supporting Information). For each dataset, a 60-images sequence is acquired with an effective framerate set to $60$ images per waveperiod (to capture exactly one wave oscillation). These relatively high effective framerates are reached by stroboscopy (the actual acquisition rate is larger than the waveperiod). The video data is then processed with the DIC algorithm~}\cite{DIC_Sander}\textcolor{\colornewtext}{~which renders $60\times2$ ($u_1$ and $u_2$) displacement maps for each frequency.}

\textcolor{\colornewtext}{\textbf{Post-processing.} Retrieving the dispersion curves requires further processing. First, the monochromatic displacement maps are converted to a single complex map by computing a discrete time-domain Fourier transform. The data is then projected on its symmetrical and anti-symmetrical as a preliminary step to the SVD operation (details of the SVD are available in Supporting Information). After selecting the relevant singular vectors, the spatial frequencies are extracted by Fourier transformation.}

}

\subsection*{\textcolor{\colornewtext}{Experimental configuration}}
\textcolor{\colornewtext}{To start off, a thin strip of dimensions $L\times w\times d=600\textrm{ mm}\times39\textrm{ mm}\times3\textrm{ mm}$ is  prepared in a soft silicone elastomer (for details see section Materials and Methods) and seeded with dark pigments for tracking purposes.
The strip is then suspended and connected to a point-like source consisting of a clamp mounted on a low-frequency (1~Hz to 200~Hz) shaker. When vibrated, the strip hosts the propagation of guided elastic waves travelling along the vertical direction~$x_1$ (see Fig.~\ref{fig_1}). The lower end of the strip is immersed in glycerol to avoid spurious reflections as well as out-of plane motions. Here, we specifically study in-plane motions \textit{i.e.} displacement components $u_1$ and $u_2$ corresponding to respective directions $x_1$ and $x_2$. The low-frequency regime enables the optical monitoring of the in plane motion. A 60 images sequence corresponding to a single wave period is acquired thanks to stroboscopic means before being processed with a Digital Image Correlation (DIC) algorithm~\cite{DIC_Sander} which retrieves the displacement of the dark seeds. Typical displacement fields ($u_1$, $u_2$) measured when shaking at 110~Hz are reported on Fig.~\ref{fig_2}(a). This method is sensitive to displacement magnitudes in the micrometer range and thus enables field extraction to be performed over large areas in spite of the significant viscous damping.\\}

\subsection*{Free edges configuration}

The interpretation of the displacement maps is not straightforward. As for any wave-guiding process the field gathers contributions from several modes. Given the system geometry, we project the data on their symmetrical (resp. anti-symmetrical) component with respect to the vertical central axis\ifthenelse{{\x=1}}{\textcolor{\colorremovedtext}{\st{(dotted line in Fig.~2)}}}{}. 
For improved extraction performances, a single value decomposition \textcolor{\colornewtext}{(SVD)} is then \ifthenelse{{\x=1}}{\textcolor{\colorremovedtext}{\st{ performed on the projected field and only the main components are kept}}}{}\textcolor{\colornewtext}{ operated and only the significant solutions are kept (for details see the Supporting Information). }\ifthenelse{{\x=1}}{\textcolor{\colorremovedtext}{\st{As depicted in Fig.}\ref{fig_2}\st{, at 110~Hz, the raw data decomposes onto}}}{}\textcolor{\colornewtext}{For example, at 110~Hz, the raw data (see Fig.~\ref{fig_2}) gathers} three main contributions: two anti-symmetrical modes (denoted $A_0$ and $A_1$) and one symmetrical mode (\ifthenelse{{\x=1}}{\textcolor{\colorremovedtext}{\st{denoted}}}{}$S_0$). Each \ifthenelse{{\x=1}}{\textcolor{\colorremovedtext}{\st{of them clearly exhibits its own}}}{}\textcolor{\colornewtext}{mode goes along with a single } spatial frequency $k$ \ifthenelse{{\x=1}}{\textcolor{\colorremovedtext}{\st{in the $x_1$ direction}}}{}\textcolor{\colornewtext}{which we extract by Fourier-transforming the right-singular vectors (containing the information relative to the $x_1$ direction)}\ifthenelse{{\x=1}}{\textcolor{\colorremovedtext}{\st{; a value extracted by performing a simple Fourier analysis of the right pseudo-eigenvector after the singular value decomposition}}}{}. \ifthenelse{{\x=1}}{\textcolor{\colorremovedtext}{\st{The procedure is repeated from 1 to 200~Hz to obtain}}}{}\textcolor{\colornewtext}{Repeating this procedure for frequencies ranging from 1 to 200~Hz, one obtains} the full dispersion diagram displayed in Fig.\ref{fig_2}(c) (filled symbols \ifthenelse{{\x=1}}{\textcolor{\colorremovedtext}{\st{are the extracted ones,}}}{}\textcolor{\colornewtext}{correspond to values directly extracted from the data, while} empty ones are obtained by symmetry with respect to the $k=0$ axis). \ifthenelse{{\x=1}}{\textcolor{\colorremovedtext}{\st{Several branches with}}}{}\textcolor{\colornewtext}{The dispersion diagram reveals several branches with} different symmetries and behaviors\ifthenelse{{\x=1}}{\textcolor{\colorremovedtext}{\st{ can be distinguished}}}{}. \textcolor{\colornewtext}{Here, the branches are}  \ifthenelse{{\x=1}}{\textcolor{\colorremovedtext}{\st{they are numbered}}}{}\textcolor{\colornewtext}{indexed} with increasing cut-off frequencies. \ifthenelse{{\x=1}}{\textcolor{\colorremovedtext}{\st{Note that the Fourier analysis only provides the real part of the wave number $k$. In fact, due to viscous dissipation, this quantity is complex. As a matter of fact, this is well pictured by the decaying character of the field maps.}}}{}\textcolor{\colornewtext}{Note that, due to viscous dissipation, the wave-number $k$ is intrinsically complex valued. As a matter of fact, this is well pictured by the decaying character of the field maps (Fig~\ref{fig_2}). The Fourier analysis yields its real part (peaks location) but also its imaginary part (peaks width) which is provided in Fig.~S4 (Supporting Information).}

Those experimental results are in good agreement with theoretical predictions (solid line) obtained with a simplified model \textcolor{\colornewtext}{and by numerical simulation (both are presented in Supporting Information)}\ifthenelse{{\x=1}}{\textcolor{\colorremovedtext}{\st{(and with 3-D simulations presented in Supporting Information)}}}{}. Indeed, one can show that the in-plane modes of a given strip are analogous to the Lamb waves propagating in a virtual 2-D plate of appropriate effective mechanical properties~\cite{laurent2020}. When the strip is made of a soft material, the analogy holds for a plate fo thickness $w$, with a shear wave velocity of $v_T$, a longitudinal velocity of exactly $2v_T$. Strikingly, this amounts to acknowledging that, for a thin strip of soft material, the low frequency in-plane guided waves are independent of the bulk modulus (or equivalently of the longitudinal wave velocity) and of the strip thickness $d$. One can then retrieve the full dispersion solely from the knowledge of the strip's shear modulus~$G$, width $w$ and density $\rho$. Of course, the intrinsic dispersive properties of the soft material as well as its lossy character must be taken into account. A simple and commonly accepted model for describing the low frequency rheology of silicone polymers is the fractional Kelvin-Voigt model~\cite{meral2009surface,kearney2015dynamic,rolley2019flexible}, for which the complex shear modulus writes $G=G_0\big[1+(\text{i}\omega\tau)^n\big]$. This formalism being injected in the 2-D model, our measurements are convincingly adjusted (solid lines in Fig.\ref{fig_2}) when the following set of parameters is input: $G_0=26~\textrm{kPa}$, $\tau=260~\textrm{$\mu$s}$ and $n=0.33$. Note that this choice of parameters turns out to match relatively well the measurements obtained with a traditional rheometer (see details in Supporting Information). The transparency of the theoretical line represents the weight of the imaginary part of the wave-number $k$ \textcolor{\colornewtext}{(detailed on Fig.~S5)}. When $k$ becomes essentially imaginary, the solution is evanescent which explains why it cannot be extracted from the experiment.

Let us now comment on a few interesting features of this dispersion diagram. 
First, at low frequencies, the single symmetrical branch (labelled $S_0$) presents a linear slope, hence defining a non-dispersive propagation or equivalently a propagation at constant wave velocity. Experimentally, the latter corresponds to $\sqrt{3}v_T$ which confirms the prediction from~\cite{laurent2020}. This is somehow counter-intuitive: the displacement of $S_0$ is quasi-exclusively polarized along the $x_1$ direction, giving it the aspect of a pseudo-longitudinal wave, but it propagates at a speed independent of the longitudinal velocity. At 150 Hz, two branches cross linearly in the $k\to 0$ limit. This is the signature of a Dirac-like cone~\cite{huang2011dirac,maznev2014dirac,stobbe2017conical}. \textcolor{\colornewtext}{It is worth mentioning that, despite the 3-D character of the system, the propagation only occurs in one direction~($x_1$) which means that the cone should be regarded as a linear crossing. \textcolor{\colornewtext}{ Its slope (group velocity) is found to be $\pm2 v_T/\pi$ (see calculation in Supporting Information).} The cone, which turns out to be well} \ifthenelse{{\x=1}}{\textcolor{\colorremovedtext}{\st{Note how well it is}}}{}defined in spite of the significant damping, directly results from the incompressible nature of the soft elastomer. \ifthenelse{{\x=1}}{\textcolor{\colorremovedtext}{\st{ present in the system. Here, the cone is the result of the incompressible nature of the soft elastomer}}}{}Indeed, \ifthenelse{{\x=1}}{\textcolor{\colorremovedtext}{\st{ when $v_L\gg v_T$ (or $\nu\approx1/2$), the second and third cut-off frequencies automatically coincide}}}{}\textcolor{\colornewtext}{ the condition $v_L\gg v_T$ (\textit{i.e.} $\nu\approx1/2$) automatically yields the coincidence of the second and third cut-off frequencies}~\cite{laurent2020}. In other words, any thin soft strip would display such a Dirac-like cone. Because the cone is located \ifthenelse{{\x=1}}{\textcolor{\colorremovedtext}{\st{in}}}{}\textcolor{\colornewtext}{at} $k=0$, the lower frequency part of the $S_2$ branch features negative wave numbers (solid symbols). \ifthenelse{{\x=1}}{\textcolor{\colorremovedtext}{\st{This means that}}}{}\textcolor{\colornewtext}{In this region,} the phase and group velocities are anti-parallel~\cite{bramhavar2011negative,philippe2015focusing}\ifthenelse{{\x=1}}{\textcolor{\colorremovedtext}{\st{and results in backward propagation of the wave-fronts (see video in Supporting Information)}}}{}. \textcolor{\colornewtext}{More specifically, the group velocity remains positive (as imposed by causality) when the phase velocity becomes negative \textit{i.e.} the wave-fronts travel toward the source (see video S3). This effect has been the scope of many developments in the metamaterials field~\cite{deymier,BookMetamaterials2} but occurs spontaneously here.}\ifthenelse{{\x=1}}{\textcolor{\colorremovedtext}{\st{It is interesting to note that the wave number extraction from experimental maps provides negative values (filled symbols) which evidences a negative phase velocity, a feature that has been the scope of many developments in the metamaterials field}~\cite{deymier,BookMetamaterials2}\st{, and that occurs spontaneously here.}}}{} 

\subsection*{Fixed edges configuration}
From now on, we implement Dirichlet boundary conditions on a $w=50.6$~mm strip by clamping its edges in a stiff aluminium frame (video S4). \ifthenelse{{\x=1}}{\textcolor{\colorremovedtext}{\st{The previous experimental steps are repeated and the dispersion curves extracted~(Fig.}~\ref{fig_3}\st{)}}}{}\textcolor{\colornewtext}{Again, the dispersion curves~(Fig.~\ref{fig_3}) are extracted following the previous experimental steps}. \textcolor{\colornewtext}{See how the low order branches ($A_0$ and $S_0$ in Fig.~\ref{fig_2}(b)) have disappeared as a consequence of the field cancellation at the boundaries. Besides, a } Dirac-like cone is observed for this configuration as well \textcolor{\colornewtext}{but it now occurs at the crossing of anti-symmetrical branches}. \textcolor{\colornewtext}{Just like in the free edges configuration, the slope at the Dirac point is $v_g=\pm2 v_T/\pi$. Extracting the field patterns for this particular point, one finds that the motion is elliptical (video S5). The polarization even becomes circular at a distance $\pm w/6$ from the centre of the strip. All these observations are supported by the calculation provided in Supporting Information. 
} \ifthenelse{{\x=1}}{\textcolor{\colorremovedtext}{\st{However, the crossing branches are now the anti-symmetrical ones. Additionally, due to the prescribed displacements at the boundaries, the lower order branches $A_0$ and $S_0$ (from Fig.}~\ref{fig_2}\st{(b)) have been suppressed and the first cut-off frequency is now associated with a symmetrical mode.}}}{} Once again, the prediction obtained with \ifthenelse{{\x=1}}{\textcolor{\colorremovedtext}{\st{a}}}{}\textcolor{\colornewtext}{the} 2-D \textcolor{\colornewtext}{equivalence} model assuming rigid boundaries convincingly matches the experiment. \ifthenelse{{\x=1}}{\textcolor{\colorremovedtext}{\st{It is also worth noticing how the slope of the $A_1$ branch switches sign for a non-zero wave number. As a consequence, the group velocity vanishes for a given frequency. This is the signature of a ZGV point; a phenomenon which has been previously observed in rigid plates}~\cite{holland2003air,prada2005laser,prada2008local,Yantchev2011,Ces2011}.\st{
In this region, the experiment and the prediction disagree. It should be noted that the latter corresponds to a free-space (no source) scenario and thus provides an estimate of the so-called \emph{real group velocity}}~\cite{muschietti1993real}\st{. Instead, the experiment is performed with a source which continuously feeds the strip. In consequence, it rather renders the \emph{energy velocity}}~\cite{simonetti2005meaning}\st{. Identical for propagating modes, these two quantities bifurcate as the imaginary part of $k$ becomes significant.}}}{}\textcolor{\colornewtext}{Also, an interesting feature shows up at 102~Hz where the branches $A_1$ and $A_2*$ nearly meet each-other. In a non-dissipative system, one expects the two branches to connect thus yielding a singular point associated with a Zero Group Velocity (ZGV); a phenomenon which has been previously observed in rigid plates~\cite{holland2003air,prada2005laser,prada2008local,prada_power_2008,grunsteidl_inverse_2016}. Here, because the propagation is damped by viscous mechanisms, the connection does not strictly occur, the reason why we talk about pseudo-ZGV mode, but as we will see below similar wave phenomena still exist in the presence of damping (see Fig. S2 for an analytical comparison between the conservative and dissipative scenarii).}\\

Let us now illustrate the rich physics associated to this dispersion diagram by \ifthenelse{{\x=1}}{\textcolor{\colorremovedtext}{\st{selectively exciting}}}{}\textcolor{\colornewtext}{specifically selecting} a few \textcolor{\colornewtext}{interesting} modes (videos S6 to S9). To begin with, the source is placed in the \ifthenelse{{\x=1}}{\textcolor{\colorremovedtext}{\st{ placed in the middle of the strip}}}{}\textcolor{\colornewtext}{centered} and shaken vertically at 136~Hz. This excitation is intrinsically symmetrical and only $S_1$ should be fed at this frequency. The chronophotographic sequence displayed on Fig.~\ref{fig_4}(a) reports twelve successive snapshots of the displacement $u_1$ taken over a full period of vibration at $136~$Hz. As expected, the field pattern respects the $S_1$ symmetry. Also, the zeroes of the field (red dashed lines) move away from the source, which corresponds to diverging waves.

On either side of the strip, there are two solutions with identical profiles but opposite phase velocities; in other words \ifthenelse{{\x=1}}{\textcolor{\colorremovedtext}{\st{they are}}}{}\textcolor{\colornewtext}{two} time-reversed partners. Thus, the bottom part of the strip hosts the solution $S_1$ while its top part supports $S_1^*$. Furthermore, the transverse field $u_2$ is \ifthenelse{{\x=1}}{\textcolor{\colorremovedtext}{\st{in phase quadrature}}}{}\textcolor{\colornewtext}{$\pi/2$ phase shifted} compared to $u_1$ at this frequency (see Fig. S7 or video S6). This essentially suggests that the in-plane displacement is \ifthenelse{{\x=1}}{\textcolor{\colorremovedtext}{\st{circularly}}}{}\textcolor{\colornewtext}{elliptically} polarized; an interesting feature since such a polarization is known to flip under a time-reversal operation. One can easily take advantage of this effect by imposing a chiral excitation.  
To this end, we use a source made of two counter-rotating clamps \textcolor{\colornewtext}{located }at equal distances from the centre of the strip\ifthenelse{{\x=1}}{\textcolor{\colorremovedtext}{\st{ (this is realised by driving two distinct clamps with 4 different speakers connected to a Presonus AudioBox 44VSL soundboard)}}}{}\textcolor{\colornewtext}{. The rotating motion is produced by driving two distinct clamps with 4 different speakers connected to a soundboard (Presonus AudioBox 44VSL)}. As depicted in Fig.~\ref{fig_4}(b), such a chiral source excites the $S_1$ mode which propagates towards $x_1>0$, however, it cannot produce its time reverse partner $S_1^*$ propagating in the opposite direction. By controlling the source's chirality, we performed selective feeding and one-way wave transport, a feature which has recently \textcolor{\colornewtext}{been exploited in different contexts~\cite{shi_observation_2019,long_symmetry_2020,yves_structure-composition_2020}}\ifthenelse{{\x=1}}{\textcolor{\colorremovedtext}{\st{stimulated the research of classical analogues to the quantum hall effect}~\cite{Soljacic,wang2015topological,mousavi2015topologically}}}{}.  

One can also try to capture the strip behaviour near the \textcolor{\colornewtext}{pseudo} ZGV point. As it is associated with an anti-symmetrical motion, the system is shaken horizontally by two clamps driven simultaneously at $102~$Hz, and the field displacement $u_2$ over a full cycle is represented in Fig.~\ref{fig_4}(c).
\ifthenelse{{\x=1}}{\textcolor{\colorremovedtext}{\st{It exhibits a very unique property: still standing zeroes denote the stationary character of the solution.}}}{}\textcolor{\colornewtext}{It exhibits a very unique property: the zeroes remain still (see dashed lines) whatever the phase within the cycle which indicates that the solution is stationary.}
\ifthenelse{{\x=1}}{\textcolor{\colorremovedtext}{\st{This feature is a classical signature of ZGV points.}}}{} To understand \ifthenelse{{\x=1}}{\textcolor{\colorremovedtext}{\st{it}}}{}\textcolor{\colornewtext}{this feature}, let us take a look back at Fig.~\ref{fig_3}. Causality imposes that $A_1$ and $A_2$ (filled symbols, solid lines) propagate in the bottom part of the strip while their time partners $A_1^*$ and $A_2^*$ (empty symbols, dashed lines) travel toward the top part. Interestingly, at \ifthenelse{{\x=1}}{\textcolor{\colorremovedtext}{\st{the ZGV frequency (102~Hz)}}}{}\textcolor{\colornewtext}{102 Hz}, $A_1$ and $A_2$ (resp. $A_1^*$ and $A_2^*$) have almost opposite wave numbers and interfere \ifthenelse{{\x=1}}{\textcolor{\colorremovedtext}{\st{in a}}}{}\textcolor{\colornewtext}{to produce a} standing wave. \ifthenelse{{\x=1}}{\textcolor{\colorremovedtext}{\st{The principle is comparable to that of a vibrating string with fixed terminations for example. Yet, in the present case, the backward wave does not result from edge's reflections but rather from the dispersion itself.}}}{}\textcolor{\colornewtext}{The stationarity does not result from some reflection at the strip ends but is a direct consequence of the coincidence of the two branches. In our damped case where the exact coincidence seems lost, the difference in magnitudes between the respective wavenumbers is sufficiently small to guarantee this effect at the pseudo-ZGV frequency.
}

Again, introducing some chirality will result in breaking the time-reversal symmetry. The sources are now rotated in an anti-symmetrical manner (see inset) resulting in the measurements reported on Fig.~\ref{fig_4}(d). The propagative nature of the field is retrieved on both sides: the zeroes of the field are travelling. Note that, on the upper part, the wave-fronts are anti-causal, \textit{i.e.} they seem to move towards the source which is typical of a negative phase velocity. 
Strictly speaking, only $A_1$ (resp $A_2^*$) remains in the lower part (resp. upper part) of the strip.
Thanks to the chiral excitation, we have separated the two contributions of a \textcolor{\colornewtext}{pseudo-}ZGV point, and highlight their unique nature as a superposition of two modes propagating in opposite directions.

\subsection*{Perspectives}

In this article, we report the observation of Dirac-like cones in a soft material in spite of a significant dissipation due to viscous effects. The associated dispersion is also found to induce atypical wave phenomena such as a negative phase velocity and a stationnary mode. For both the Dirichlet and Neumann boundaries, a convincing agreement is found between experiments, the theoretical simplified model and numerical analysis. \ifthenelse{{\x=1}}{\textcolor{\colorremovedtext}{\st{Additionally, we perform the selective feeding of chiral waves into the strip.}}}{}\textcolor{\colornewtext}{Additionally, we perform selective feeding by controlling the chirality of the source.}\ifthenelse{{\x=1}}{\textcolor{\colorremovedtext}{\st{These experimental results show how the physics of semi-conductors can echo in a drastically different context.}}}{} \ifthenelse{{\x=1}}{\textcolor{\colorremovedtext}{\st{Beyond the wave physics phenomena}}}{}\textcolor{\colornewtext}{Beyond the original wave physics}, the soft strip \ifthenelse{{\x=1}}{\textcolor{\colorremovedtext}{\st{case}}}{}\textcolor{\colornewtext}{configuration} may stimulate interest in different domains in a near future. From a material point of view, we show how a very simple platform can provide comprehensive information about the visco-elastic properties of a soft solid leading to new technologies to probe \ifthenelse{{\x=1}}{\textcolor{\colorremovedtext}{\st{the}}}{}\textcolor{\colornewtext}{its} rheology. From a biological point of view, understanding the complex physics associated with a geometry that is ubiquitous in the human tissues and organs, is a major challenge. Imaging and therapeutic methods based on elastography \ifthenelse{{\x=1}}{\textcolor{\colorremovedtext}{\st{could}}}{}\textcolor{\colornewtext}{would} benefit from an in-depth understanding of the specific dynamic response of tendons~\cite{brum2014vivo}, myocardium~\cite{nenadic2011lamb} or vocal cords~\cite{ishizaka1976computer} among others. Some physiological mechanisms could also be unveiled by accounting for the atypical vibrations of a soft strip. In the inner ear, for instance, the sound transduction is essentially driven by a combination of two soft strips namely the basilar and tectorial membranes~\cite{reichenbach2014,sellon2015longitudinal,sellon2019nanoscale}. Overall, we might soon discover that evolution had long transposed the exceptional properties of graphene to the living world.

\showmatmethods{} 

\acknow{We thank Sander Wildeman for sharing his DIC algorithm, Gatien Cl{\'e}ment for early experimental developments, J{\'e}r{\^o}me Laurent for sharing his Lamb modes dispersion code and Pascale Arnaud for video editing. This work has been supported by LABEX WIFI (Laboratory of Excellence within the French Program “Investments for the Future”) under references ANR-10-LABX-24 and ANR-10-IDEX-0001-02 PSL*, and by Agence Nationale de la Recherche under reference ANR-16-CE31-0015.}

\showacknow{} 

\bibliography{biblio}

\end{document}



\maketitle

\SItext

\section*{From bulk waves to in-plane guided waves in thin rectangular beam.}
Bars of rectangular cross-section are  elastic wave guides. As three wave polarizations are coupled in this geometry, finding the full dispersion diagram can reveal complex~\cite{krushynska_normal_2011}. Fortunately, for the thin strip studied in this paper, the problem drastically simplifies thanks to a 2-D analogy. This analogy is detailed in reference~\cite{laurent2020}. Here, we recall its key features.\\

\begin{figure}
    \begin{center}
        \includegraphics[width = \textwidth]{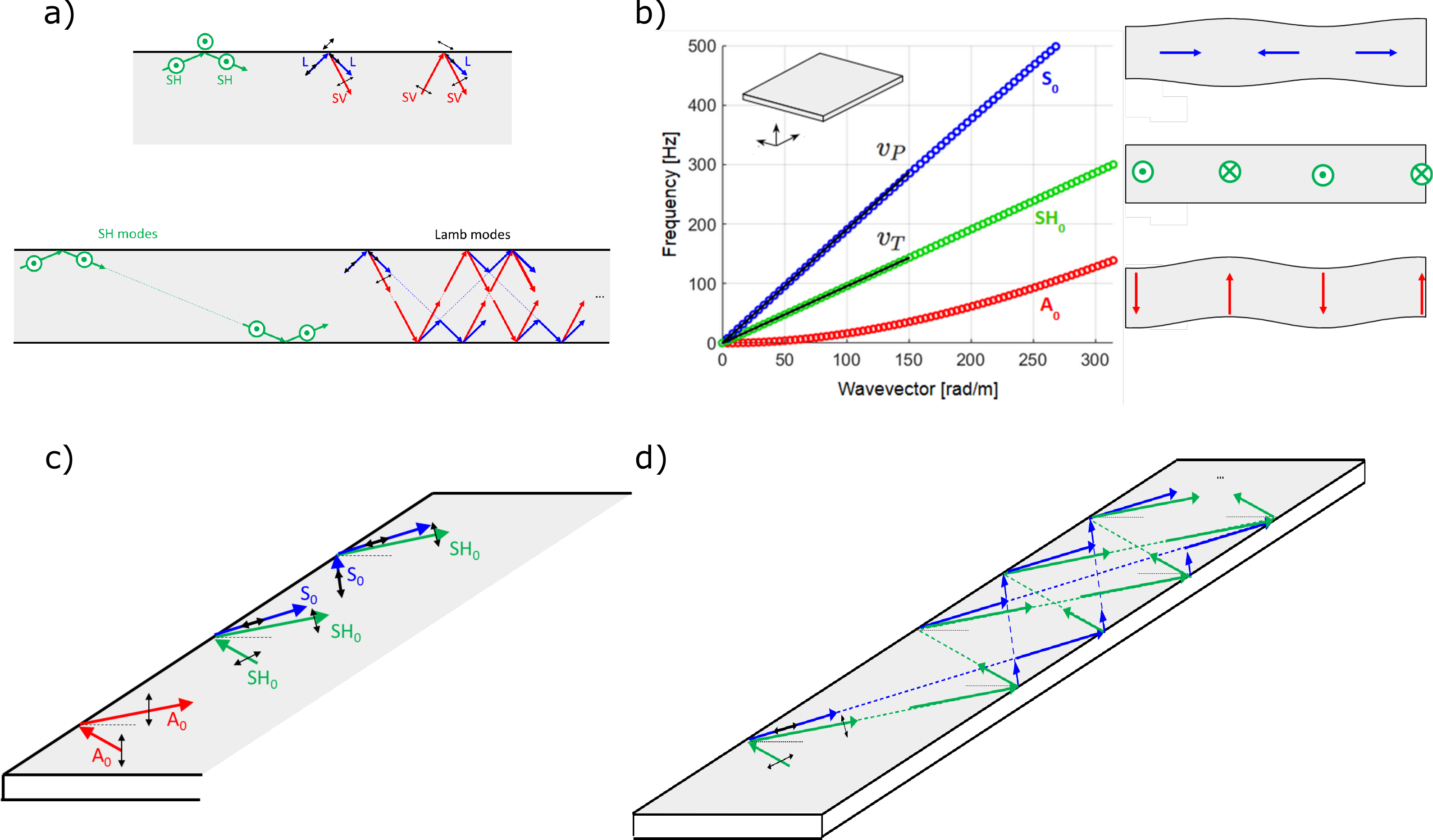}
        \caption{\textbf{Theoretical considerations.} (a) Illustration of the coupling between bulk waves at each reflection: $SV$ and longitudinal waves are coupled  while the $SH$ ones remain uncoupled. It gives rise to 2 families of guided modes in a plate. (b) Low frequency dispersion relation for a 3-mm-thick-plate made of nearly incompressible material with $v_T=6 \textrm{m.s}^{-1}$. In this low frequency limit each branch can be associated to its own pseudo-polarization: $S_0$ looks like a longitudinal wave, $SH_0$ is transversely polarized shear wave, and $A_0$ looks like a vertically polarized transverse wave. Interestingly the $S_0$ travels twice faster than $SH_0$. (c) Adding a boundary to the plate, the analogy with the coupling of (a) can be made: $S_0$ and $SH_0$ are now coupled at each reflection while $A_0$ remains alone. (d) Illustration between the analogy of the low frequency in-plane waves in a thin strip and the Lamb waves in the plate of (a).}
        \label{FigureSI_Theory}
    \end{center}
\end{figure}

Bulk elastic waves propagating in isotropic materials can be decomposed on three distinct polarizations: a longitudinal polarization travelling with a velocity $v_L$ and two transverse polarizations (or shear waves) propagating at the speed $v_T$. In the presence of a horizontal interface, reflections occur and the polarisation of the so-called shear horizontal wave (SH)  remains unaffected while, the longitudinal (L) and shear vertical (SV) polarizations couple with each other (figure~\ref{FigureSI_Theory}.a). \\

By adding a second parallel interface,  the host medium becomes a plate and supports two families of independent guided modes: the SH modes resulting from multiple reflections of SH waves and the so-called Lamb modes which result from the coupling between multiply reflected SV and L waves (figure~\ref{FigureSI_Theory}.a).\\

At low frequencies, only 3 guided modes propagate in the plate (figure~\ref{FigureSI_Theory}.b): the first SH mode (SH$_0$), and the two Lamb modes which correspond respectively to a symmetric (S$_0$) and anti-symmetric (A$_0$) displacement pattern. In the low frequency limit, their profiles along the plate thickness are relatively homogeneous and they can be considered as nearly linearly polarized. In particular, S$_0$ can be seen as a pseudo-longitudinal wave propagating 
at a constant "plate velocity" $v_P$. Nearly incompressible materials ($\nu\approx 1/2$), such as the one we investigate in the main text, are particularly interesting since $v_P=2v_T$ (figure~\ref{FigureSI_Theory}.b). \\

All of these observations enable the construction of an analogy between Lamb waves in a plate and the in-plane guided waves within a thin strip as sketched in figure~\ref{FigureSI_Theory}.c. Similarly to SH modes in plates the $A_0$ mode remains independent at each reflection along the edges of the strip. However, SH$_0$ and S$_0$ behave similarly to the longitudinal and shear vertical bulk waves \textit{i.e.} they couple at each reflection. Adding a second parallel interface, we now understand (figure~\ref{FigureSI_Theory}.d) that SH$_0$ and S$_0$ give rise to the in-plane guided modes studied in the main text. \\

As a summary, the low frequency dispersion diagram for the in-plane guided waves in a strip do not depend on the strip thickness and  can be retrieved by solving the Rayleigh-Lamb equation for a virtual plate of thickness equal to the strip width, shear velocity of $v_T$ and longitudinal velocity of $v_P$. Interestingly, for incompressible materials, we have  $v_P=2v_T$ meaning that the knowledge of $v_T$ is sufficient to obtain the dispersion diagram. In other words, from the knowledge of the shear modulus $G=\rho v_T^2$ one can numerically solve the Rayleigh-Lamb problem and obtain the dispersion curves.

\section*{Dispersion  of in-plane modes for free and fixed edges}
Using the Rayleigh Lamb approximation, the general expressions of the parallel $(u_1)$ and normal $(u_2)$ displacement components for symmetrical $(\alpha=0)$ and anti-symmetrical $(\alpha=\pi/2)$  modes are given by~\cite{royer1999elastic}:
\begin{equation}
\left\{
    \begin{array}{ll}
		u_1(x_2,k)=-ikB\cos{(px_2+\alpha)} +qA\cos{(qx_2+\alpha)}\\
		\\
		u_2 (x_2,k)=-pB\sin{(px_2+\alpha)}+iAk\sin{(qx_2+\alpha)}
	\end{array}
\right.
\label{displacements}
\end{equation}
with $p^2=(\omega/v_P)^2-k^2$ and $q^2=(\omega/v_T)^2-k^2$. $A$ and $B$ are coefficients to relate.  The dispersion equation of these modes is obtained from the boundary conditions.\\

\noindent
$\rightarrow$\hspace*{.1cm}{\bf Neumann boundary conditions:} For a strip with free edges, the coefficients $A$ and $B$ satisfy the following equations:
\begin{equation}
\left\{ 
 	\begin{array}{ll}
  		(k^2-q^2)B\cos(ph+\alpha)+2ikqA\cos(qh+\alpha)= 0 \\
  		\\
    	2ikpB\sin(ph+\alpha)+(k^2-q^2)A\sin(qh+\alpha)= 0 
    \end{array}
    \right.
\label{EquCoef AB Neumann}
\end{equation}
where $h=w/2$ is half the strip width. Non trivial solutions exist if the following dispersion equation is satisfied: 
\begin{equation}
(k^2-q^2)^2\cos(ph+\alpha)\sin(qh+\alpha)=4k^2pq\cos(qh+\alpha)\sin(ph+\alpha) 
\label{DispersionLibre}
\end{equation}

\noindent
$\rightarrow$\hspace*{.1cm}{\bf Dirichlet boundary conditions}: For a clamped strip,  the  boundary conditions lead to the equations:  
\begin{equation}
  \left\{
  \begin{array}{ll}
		 -ikB\cos{(ph+\alpha)}+qA\cos{(qh+\alpha)} =0\\
		\\
		 -pB\sin{(ph+\alpha)}+iAk\sin{(qh+\alpha)}=0
	\end{array}
	\right.
   \label{EquCoef AB Dirichlet}
\end{equation}
The dispersion equation is then: 
\begin{equation}
k^2\cos{(ph+\alpha)}\sin{(qh+\alpha)} +qp\cos{(qh+\alpha)}\sin{(ph+\alpha)}=0
\label{dispersionClampé}
\end{equation}

\section*{Displacements and chiral excitation}
In the main text, we propose a mode selection method based on chiral excitation explicitly suggesting that the motion is elliptically polarized.
On a theoretical point of view, using the first equation of [\ref{EquCoef AB Neumann}] or [\ref{EquCoef AB Dirichlet}]   we  express $B$ as a function of $A$ and obtain expressions of the displacement fields for each boundary condition:\\

\noindent
$\rightarrow$\hspace*{.1cm}{\bf Neumann boundary conditions:}
\begin{equation} \left\{
 \begin{array}{ll}
 u_1(x_2,k)=&Aq\left(\frac{2ik}{k^2-q^2}\frac{\cos(qh+\alpha)}{\cos(ph+\alpha)}\cos(px_2+\alpha)+\cos(qx_2+\alpha)\right)\\
 u_2(x_2,k)=&iAk\left(\frac{2ip}{k^2-q^2}\frac{\cos(qh+\alpha)}{\cos(ph+\alpha)}\sin(px_2+\alpha)+\sin(qx_2+\alpha)\right)
	\end{array}
\right.
\end{equation}

\noindent
$\rightarrow$\hspace*{.1cm}{\bf Dirichlet boundary conditions: }
 \begin{equation}
 \left\{
 \begin{array}{ll}
 u_1(x_2,k)=&Aq\left(\cos(qx_2+\alpha)+\frac{\cos(qh+\alpha)}{\cos(ph+\alpha)}\cos(px_2+\alpha)\right)   \\
 u_2(x_2,k)=&iA\left(pq\frac{\cos(qh+\alpha)}{\cos(ph+\alpha)}\sin(px_2+\alpha)+k\sin(qx_2+\alpha)\right);
\end{array}
\right.
\end{equation}
	
Note that if $\cos(ph+\alpha)$ vanishes, similar expressions can be obtained by using the second equation in [\ref{EquCoef AB Neumann}] or [\ref{EquCoef AB Dirichlet}]. 
In the case $\omega/k>2v_T$ (i.e. $p$ and $q$ real) these formulas clearly exhibit the $\pm \pi/2$ phase shift  between $u_1$ and $u_2$. Except for specific positions where one of the two components vanishes and present a node, the particle therefore always exhibits an elliptical motion.  

To illustrate this effect, we provide the trajectories at four different positions of interest over a full phase cycle (see figure~\ref{FigureSI_Trajectories}). These measurements confirm the symmetrical character of the motion together with its elliptical polarization. Note how the rotation switches polarity on either sides of the source. The modes $S_1$ and $S_1^*$ are orthogonal. This suggests that an appropriate source polarity will select a given mode and thus propagate in a single direction. This is precisely the approach proposed in figure 4.b (main document), where the chiral excitation is incompatible with the propagation of the mode  $S_1^*$. 

\section*{Behaviour at the Dirac cones}

The dispersion $\omega(k)$ at small wave-number was thoroughly studied by Mindlin for a plate with free edges and presentend in chapter 2 of his book~\cite{Mindlin06}. In the vicinity of a cut-off frequency $f_c=\omega_c/2\pi$, the slope of the dispersion curve generally vanishes and  the dispersion law $\omega(k)$ can be developed to the second order as: 
\begin{equation}
	\omega(k) = \omega_c + D k^2 + o(k^2).
\end{equation}
This result does not hold in the strip if there is a coincidence between a shear SH and a S$_0$ compression resonance of the same symmetry. 
In these particular cases, the dispersion is linear near the cut off and the dispersion law can be developed to the first order as: 
 \begin{equation}
	\omega(k) = \omega_c + v_g k + o(k).
\end{equation}
where $v_g$ is the group velocity of the mode.\\

For soft materials,  we have  $v_P=2v_T$ so that  a Dirac cone occurs at the coincidence frequency $f=v_T/2h$ for the symmetrical modes of the free edges configuration  and for anti-symmetrical modes of the fixed edges configuration (figure~\ref{FigureSI_DispersionTheory}). In  both cases, a linear  slope is found: 
 \begin{equation}
	\omega(k) = \frac{2\pi v_T}{2h} + v_g k + o(k).
\end{equation}
For the free edges, $v_g=\frac{\pm 2v_T}{\pi}$, as given by Mindlin for a free plate~\cite{Mindlin06}.
For the fixed edges, $v_g$ can be simply derived using a Taylor expansion of the dispersion equation [\ref{dispersionClampé}] for anti-symmetrical modes~$(\alpha=\frac{\pi}{2})$ at  $f_c=v_T/2h$.
And the group velocity is also found to be $v_g=\frac{\pm 2v_T}{\pi}$, just like in the free edges case.

\subsection*{Displacements at Dirac cones}
To determine the displacement close to cut-off frequencies, we seek for a relation between $A$ and $B$ for $k \rightarrow 0$.\\
 
\noindent
$\rightarrow$\hspace*{.1cm}{\bf Neumann boundary conditions :} The Taylor expansion of  equation [\ref{EquCoef AB Neumann}] provides the simple relation $ B= -2isA$ where $s$ is the sign of the group velocity.
As a consequence, the displacements components near the Dirac cone at point $x_2$ while keeping only the leading order of the Taylor expansion are: 
%
  \begin{equation}
    \left\{ \begin{array}{ll}
		u_1(x_2,k) = A\frac{\pi}{h}\cos{(\frac{\pi}{h}x_2)} + O(k) \\
		  \\         
		u_2 (x_2,k) = -isA\frac{\pi}{h}\sin{(\frac{\pi}{2h}x_2)}+ O(k)   
	\end{array}  
	\right.
 \end{equation}
 Taking for example $x_2=h$ these developments simplify into
  \begin{equation*}
   \left\{ \begin{array}{ll}
		u_1(h,k)  &= -A \frac{\pi}{h} + O(k) \\
		\\
		u_2 (h,k) &= -isA  \frac{\pi}{h}+ O(k)
	\end{array}
		\right.
  \label{Approximated displacements}
\end{equation*}
Again, the factor $i$ between  components indicate a circular polarization, and $s$ indicates opposite rotations for forward and backward modes. \\

\noindent
$\rightarrow$\hspace*{.1cm}{\bf  Dirichlet  boundary conditions :} \\
Similarly, developing eq.~[\ref{EquCoef AB Dirichlet}] leads to  
$-ikB +\frac{\pi}{h} \frac{V_g}{V_T}hk A =0 $ or simply $B=-2isA$
and the displacements near the Dirac cone are written:
\begin{equation}
    \left\{
    \begin{array}{ll}
	u_1(x_2,k) 	=	 -A\frac{\pi}{h}\sin{(\frac{\pi}{h}x_2)}+ O(k)\\
		\\
		u_2(x_2,k) 	=	 isA\frac{\pi}{h}\cos{(\frac{\pi}{2h}x_2)}+O(k)
	\end{array}
		\right.
   \label{deplacementClampe}
\end{equation}
Just like for the free edges, the displacement components have a $\pi/2$ phase difference. However, in this case, as $\cos(\pi/6)=\sin(\pi/3)$, the circular polarisation occurs for $x_2=\pm h/3=\pm w/6$.
This is coherent with our experimental observations (see video S5). 

\section*{Influence of losses on the dispersion curves}
\begin{figure}
    \begin{center}
        \includegraphics[width = \textwidth]{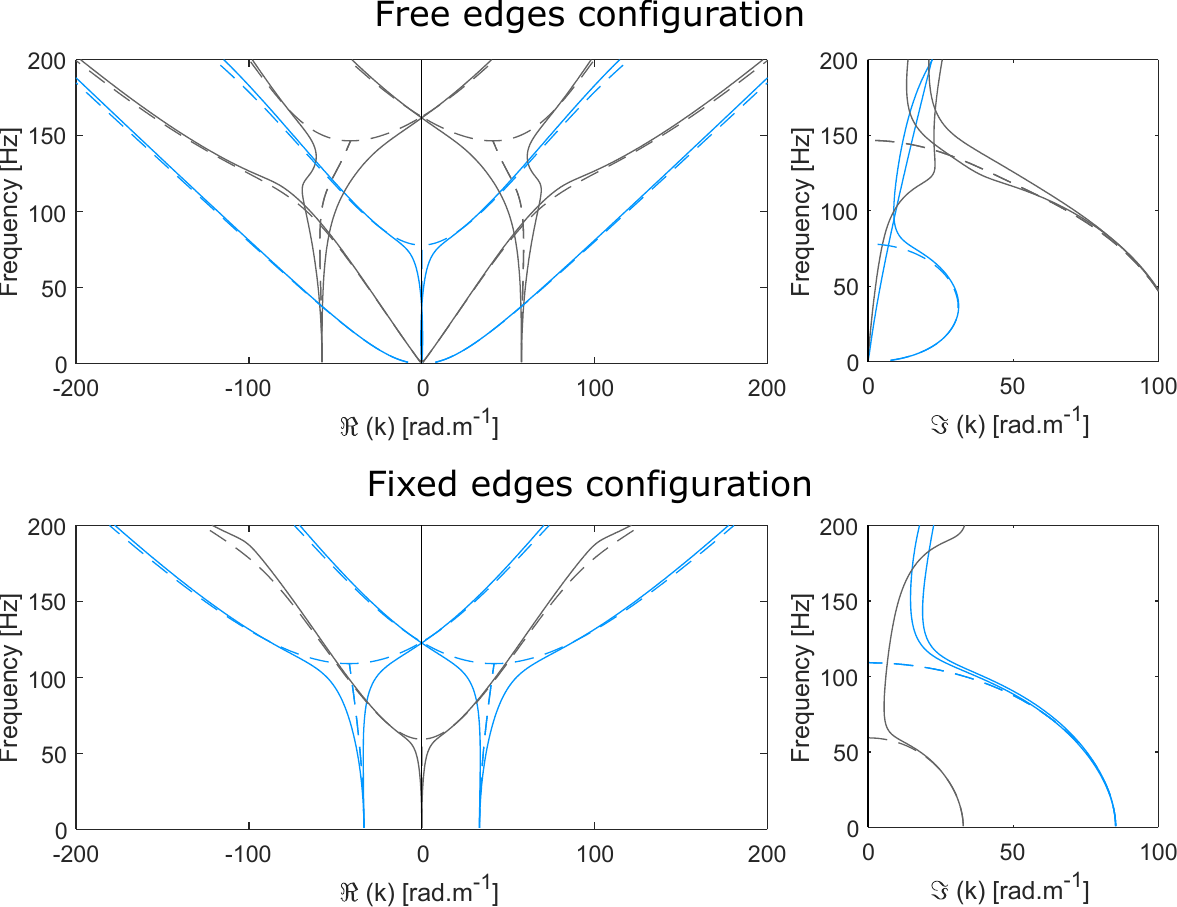}
        \caption{\textbf{Theoretical dispersion relations.} The lossless (dashed line) and lossy (continuous line) wavenumbers (real part on left and imaginary parts on right) obtained for the two configurations with the fractional Maxwell-Voigt model presented in the main text. While the ZGV point disappears when adding losses the Dirac cone remains. The imaginary parts presented here have been used for the transparency of the lines of figure 2 and 3 in the main text.}
        \label{FigureSI_DispersionTheory}
    \end{center}
\end{figure}

In all previous sections, losses are not considered. However, it has to be kept in mind that, because of viscous mechanisms, soft solids are highly dissipative systems. The shear rheology of silicone polymers is commonly described by the fractional Kelvin-Voigt model~\cite{meral2009surface,kearney2015dynamic,rolley2019flexible}, which writes $G(\omega)=G_0\big[1+(\text{i}\omega\tau)^n\big]$. For comparison with the lossless case, we only consider the the real part of the shear modulus ($G(\omega)=G_0\big[1+\cos(n\pi/2)(\omega\tau)^n\big]$). For both dispersion equations [\ref{DispersionLibre}] and [\ref{dispersionClampé}], a complex shear velocity is deduced from the shear modulus $G=\rho v_T^2$. The roots are found with a numerical Muller's method for the same set of parameters as in the main text ($G_0=26~\textrm{kPa}$, $\tau=260~\textrm{$\mu$s}$ and $n=0.33$). The real and imaginary parts of the wavenumbers are obtained for each configuration (with/without losses) and represented in figure~\ref{FigureSI_DispersionTheory}. \\

In the lossless case, we recognize the pseudo-Lamb modes described in the previous section. Notably, a clear existence of a so-called ZGV point (near$f=150$~Hz for symmetric modes in the free edges configuration, and near 110~Hz for the anti-symmetric modes in the fixed edges configuration) is evidenced. The dispersion diagram also presents the aforementioned Dirac cone at $k=0$ in the two configurations slightly above the ZGV frequency.\\

Below their cutoff frequency (the ZGV point being a simultaneous cutoff frequency for two different modes) each mode presents a solution with a non-zero imaginary part even for this lossless case. Basically it traduces the fact that evanescent waves can exist and gives access to the associated attenuation distance. Usually the branches with non-zero imaginary parts are not represented in a dispersion diagram which in turns allows to better visualize the ZGV point. This is why in the main text we decided to color-coded the theoretical lines with an increase of transparency while the imaginary part increases. \\

The addition of losses modifies the strip behavior near the first cut-off frequency and near the ZGV point for which the two branches do not connect anymore in the real plane (see figure~\ref{FigureSI_DispersionTheory} continuous lines). On the contrary, the dispersion appears unaffected at the Dirac cone which seems robust to dissipation. 
Note also that the imaginary part of the wavenumber $\Im(k)$ never exceeds 25~rad.m$^{-1}$, except for the branches below the cut-off frequencies (already present in the lossless case). 

\section*{Signal Processing}
 At a given frequency, a complex displacement field is obtained from the 60 snapshots of the strip at different time within a wave cycle thanks to the DIC procedure. The complex displacement components $u_1(\vec{r},\omega)$ along the $x_1$-direction and $u_2(\vec{r},\omega)$ along $x_2$-direction are a superposition of all possible  modes existing at this frequency that need to be identified. The two field maps are sampled on a ($N_2$,$N_1$) grid of pitch $dx$ and are then concatenated on a single complex matrix $\mathbf{U}$ of dimensions ($2N_2$,$N_1$). 
 We are then looking for the propagating modes along $x_1$-direction, thus meaning a $e^{ik x_1}$ dependence while the $x_2$ dependence remains more complex. From the matrix point of view, the single mode of wave number $k$ can be written as the product of two vectors as $D^T K$ where   $D=(D_1,....,D_{2N_2})$ is the concatenation of the complex displacements  profiles along the $x_2$-direction and $K=(1, ..., e^{ik (N_1-1) dx})$ is the spatial Fourier vector. The total displacement field can, in theory, be written as the sum of $M$ modes of wave numbers $k_m$ and complex amplitude $a_m$ and simplified in a matrix product as :
\begin{equation}
\mathbf{U} = \sum_{m=1}^M\,a_m\,D_mK_m=\mathbf{D}\mathbf{A}\mathbf{K}
\label{Decomposition}
\end{equation}
where $\mathbf{D}$ is the displacement matrix of dimensions $(2N_1,M)$,  $\mathbf{A}$ the amplitude diagonal matrix of dimensions $(M,M)$ and $\mathbf{K}$ the Fourier vector matrix of dimensions $(M,N_1)$.From this formulation, it appears that the rank of $\mathbf{U}$ is the number of modes. Furthermore, if the spatial sampling is sufficient and the wave number different, the Fourier vectors $K_m$ are  orthogonal, so that  equation (\ref{Decomposition}) is close to a singular value decomposition (SVD). Indeed, the SVD of the measured displacement field $\mathbf{U}$ is written as the product of three matrices as
\begin{equation}
\mathbf{U} = \mathbf{V} \Sigma\mathbf{W}^*
\label{eq-SVD}
\end{equation}
where $\mathbf{V}$ and $\mathbf{W}$ are unitary matrices. If $\mathbf{U}$ is an $(2N_2,N_1)$ matrix, then $\mathbf{V} $ is an $(2N_2,M)$ matrix (the left singular vectors), $\mathbf{\Sigma}$ is an $(M,M)$ diagonal real matrix containing the singular values $\lambda_m$ and $\mathbf{W}$ is an $(M,N_1)$ matrix (the right singular vectors). Because the rank of $\mathbf{U}$ is in theory equal to the number of modes $M$,  $\Sigma$ contains only  $M$  non-zero values.  In practice, if the data is good enough, there are $M$ significant singular values above noise singular values. By identification between the matrices of equation (\ref{Decomposition}) and of equation (\ref{eq-SVD}), it appears that the wave number can be obtained by Fourier transform of each right singular vector. Furthermore, although it is known that the modes profiles are not fully orthogonal, in most cases they are correctly provided by the left singular vectors. The shape of the mode number $m$ is then  given by the matrix $V_m W_m^*$. 

This method is efficient provided that different modes have different wave numbers. To avoid mode crossings, we treat separately the symmetrical and anti-symmetrical parts of the data. 

\begin{figure}
    \begin{center}
        \includegraphics[width = \textwidth]{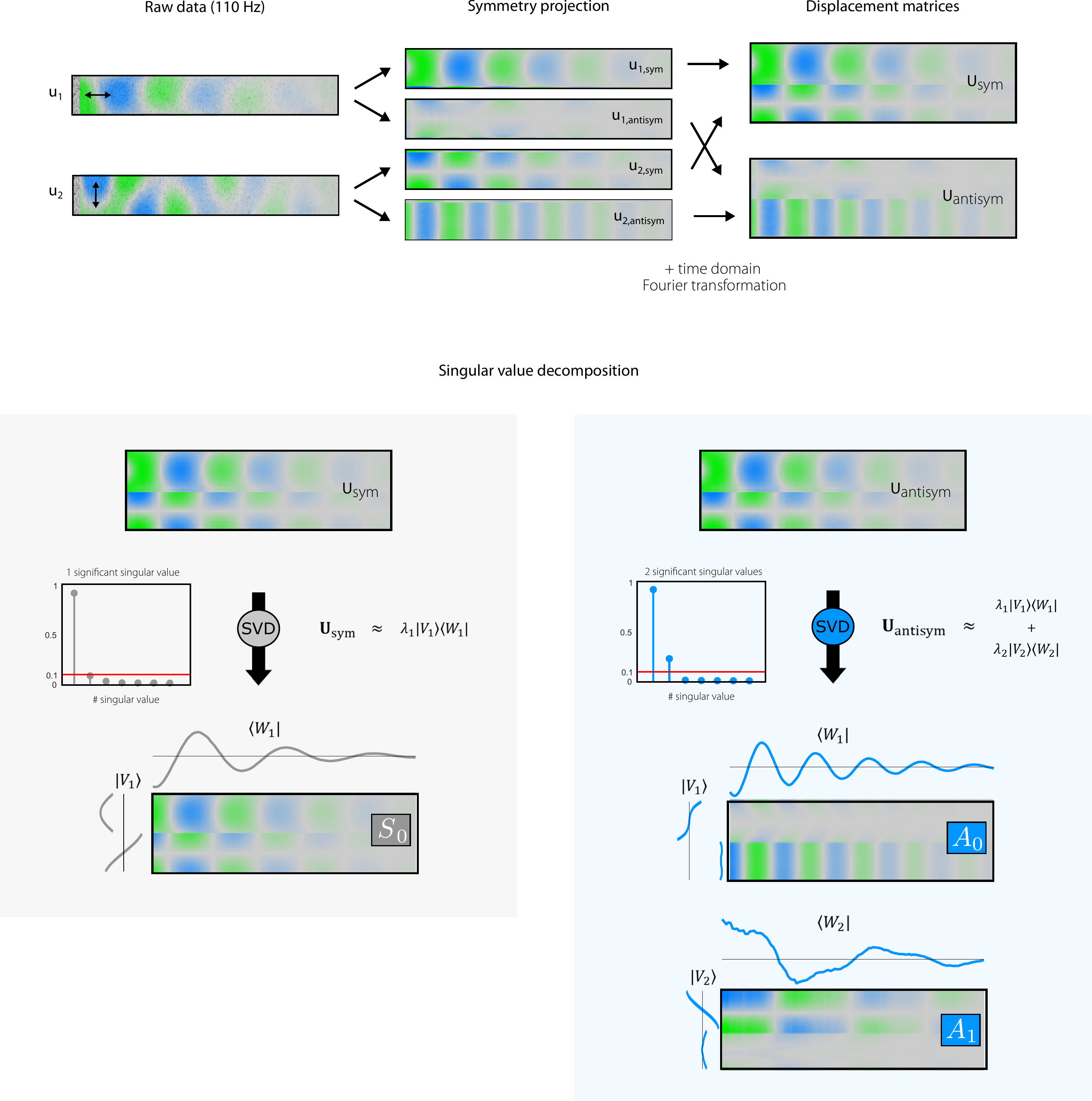}
        \caption{Description of the Singular Value Decomposition (SVD) algorithm (here with data obtained at 110~Hz, \textit{i.e.} corresponding to figure 2 in the main text). After acquiring the raw in-plane motion is acquired thanks to Digital Image Correlation (DIC) techniques, the displacement components $u_1$ and $u_2$ are projected onto their symmetrical and anti-symmetrical parts and converted to complex fields by performing a time-domain Fourier transform. The symmetrical (resp. anti-symmetrical) data are then concatenated into a single complex $ 2N_2\times N_1$ matrix $\mathbf{U}_\text{sym}$ (resp. $\mathbf{U}_\text{antisym}$) on which the SVD operation is directly applied. After extracting the most significant modes (we set the threshold at 10\% of contribution in magnitude), we obtain (at 110~Hz) one symmetrical ($S_0$) and two anti-symmetrical ($A_0$ and $A_1$) modes.}
        \label{SVD}
    \end{center}
\end{figure}

The whole procedure is illustrated in Figure \ref{SVD}. Each displacement couple $u_1,u_2$ is first decomposed onto its symmetric and anti-symmetric parts. Using the SVD, these are then decomposed into distinct modes. In this example, two singular values are above our threshold (10\% of the maximum singular value) for the anti-symmetric part and only one for the symmetric part. 
\\

\section*{Experimental dispersion curves}
\begin{figure}
\centering
\includegraphics{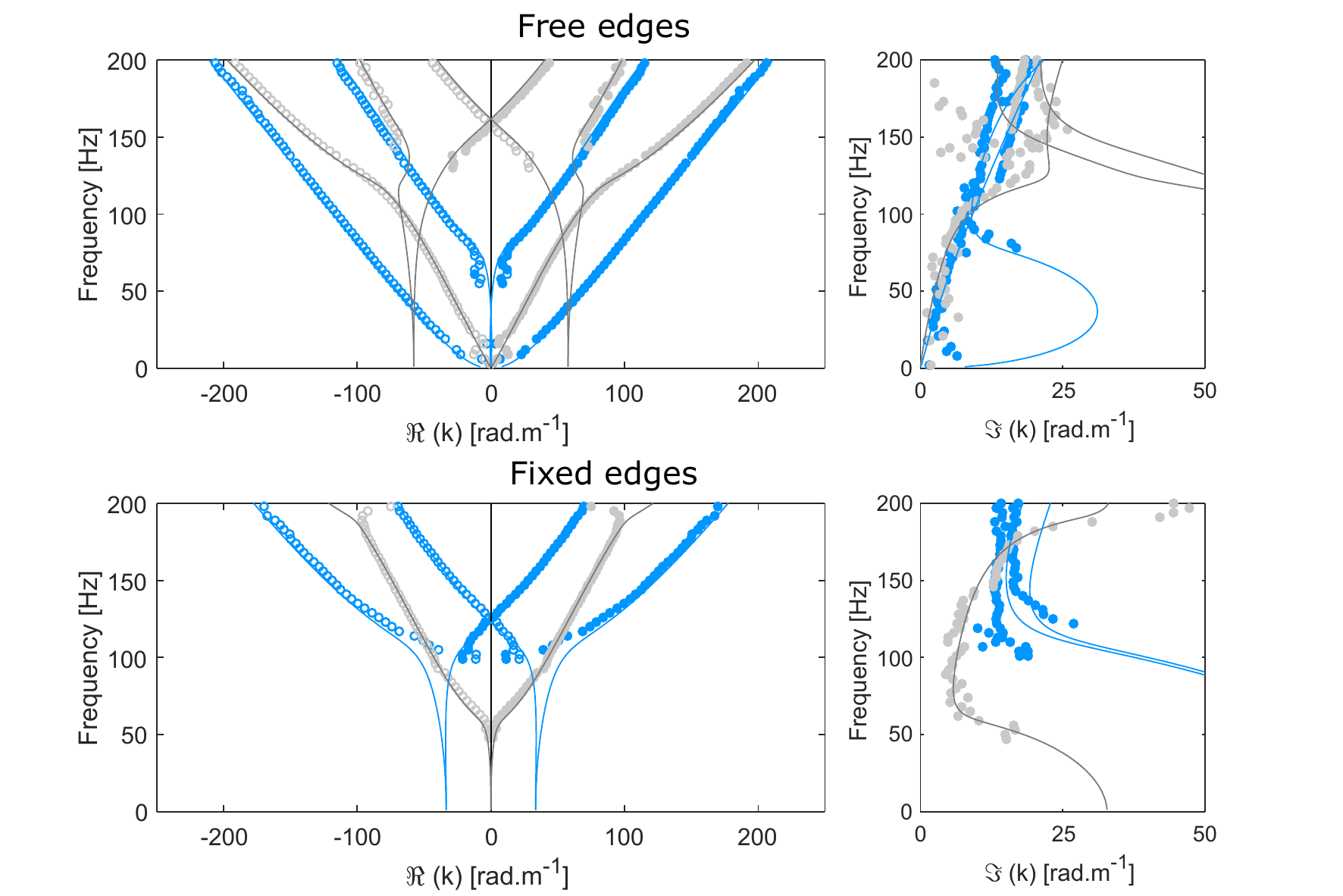}
\caption{\textbf{Experimental dispersion relations.} The real parts are the same as figure 2 and 3 of the main text, except that the theoretical line does not present transparency when the imaginary part increases. The extracted imaginary parts on the same experimental data are represented on the right.}
\label{FigureSI_DispersionExperiments}
\end{figure}

The signal processing steps described above are repeated for each frequency giving the dispersion curves of figure 2 and figure 3 (main text). On top of the real part of the wave-number presented in the main text, we also extracted its imaginary part and display the results in figure~\ref{FigureSI_DispersionExperiments}. This was performed by evaluating the width of the peak of the Fourier transform of each significant right pseudo-eigenvector obtained after SVD. When the width corresponded to the limit case of the width of our observation window, the imaginary part of the wave-number was not measured in the reciprocal space but directly estimated from an exponential fit in the real space.

Extracting the imaginary parts is always a challenging process. Here the measurements are not as accurate as for the real parts. But they still seem in relatively good agreement with the theoretical predictions. The main trends, such as the increase in the attenuation of $S_0$ at 100~Hz (fig. S4) are well captured. 

\section*{Fast shear rheology}
\begin{figure}
\centering
\includegraphics[width=11.6cm]{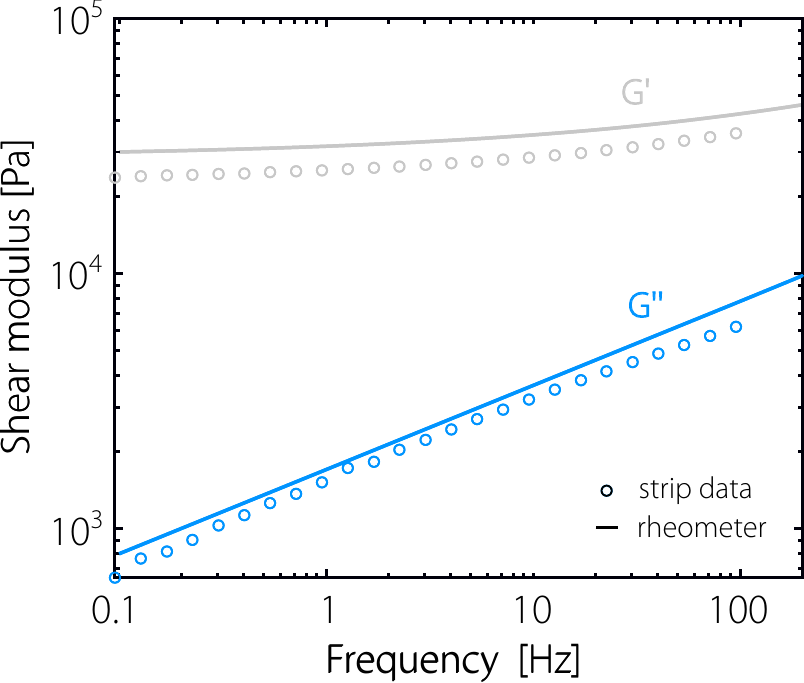}
\caption{\textbf{Shear rheology.} Shear modulus ($G=G'+\text{i}G''$) as a function of frequency. Law extracted from the strip experimental data (lines) and measurements with a Plate-Plate rheometer (symbols).}
\label{fig_1si}
\end{figure}

As discussed in the theoretical part the 2-D approximation efficiently renders our experimental data. To confirm the values obtained for $G$, we performed shear modulus measurements using a conventional rheometer (Anton-Paar MCR501) in plate-plate configuration. We measure the shear modulus $G(\omega) = G^\prime(\omega) + \text{i} G^{\prime\prime}(\omega)$ for frequencies ranging from 0.1 to 100 Hz and report the results in figure \ref{fig_1si} (open symbols). In the probed range, both $G^\prime$ and $G^{\prime\prime}$ convincingly agree with the fractional Kelvin-Voigt model extracted from strip experiment. However, the strip method seems to overestimate the storage modulus $G^{\prime}$ by roughly~$10\%$. We believe that this is related to slightly different curing conditions in the two experiments. Note, that the frequency range one can access with the strip experiment is twice larger than that of the shear rheometer. The main limitation being that high frequency modes are more attenuated and thus harder to probe with optical means. Sensibly higher frequencies would be made available by simply using a better shaker or a magnifying lens. 

\section*{Numerical simulation}
\begin{figure}
\centering
\includegraphics{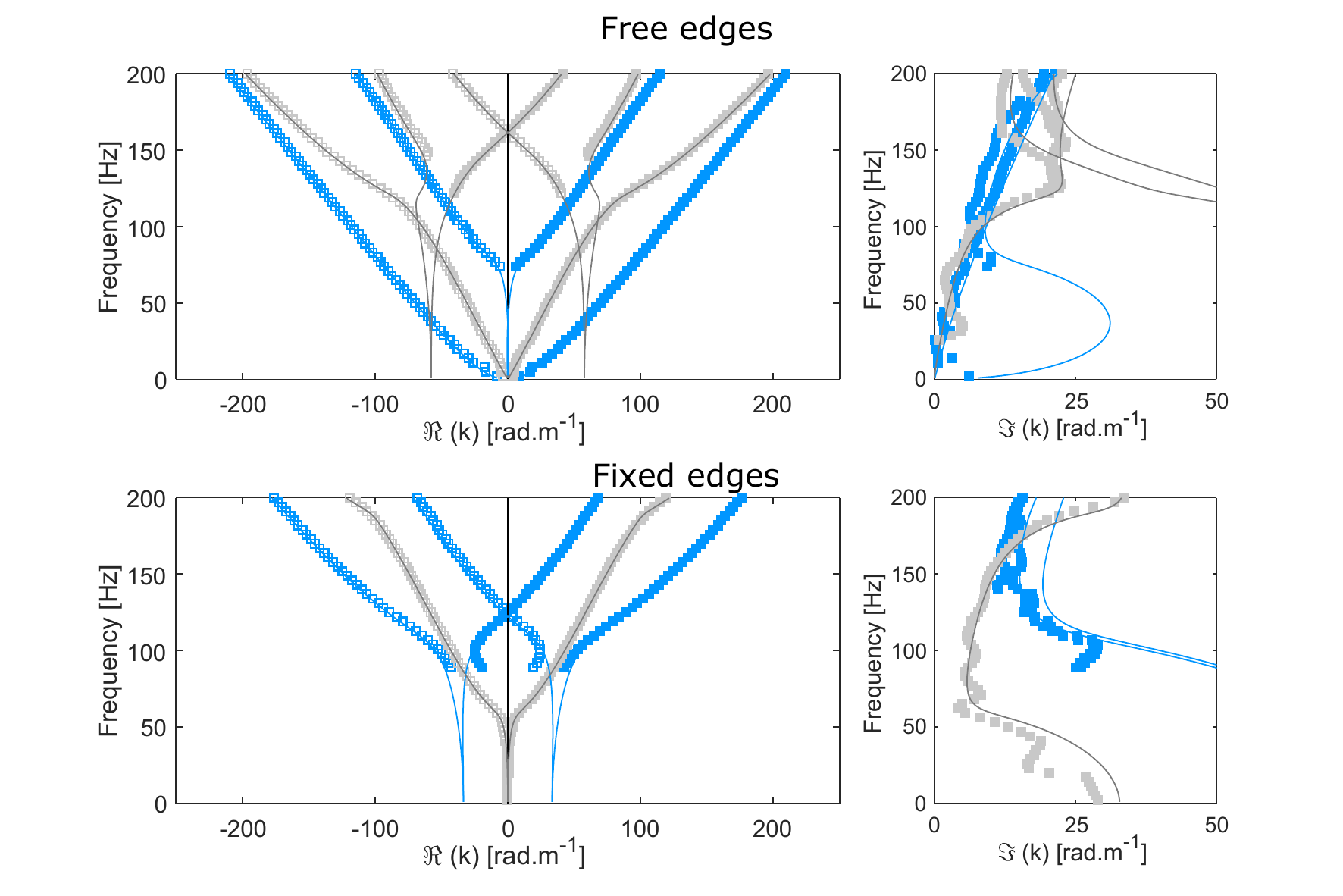}
\caption{\textbf{Finite element simulation.} Same as figure~\ref{FigureSI_DispersionExperiments} but on data obtained by numerical simulation on a full 3-D strip with the same mechanical properties.}
\label{FigureSI_Simulations}
\end{figure}

\begin{figure}
    \begin{center}
        \includegraphics[width = 15cm]{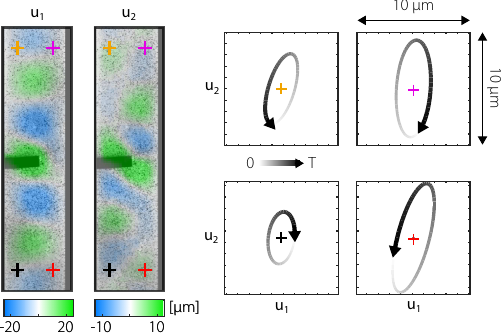}
        \caption{\textbf{Fixed edges field patterns} (left) Snapshot of the field $u_1$ (same data as figure 4 a) and $u_2$ for a vertical excitation on the fixed edges strip. Represented over a full cycle the displacement measured on four different points exhibit elliptical displacements responsible for the selective excitations presented in figure 4. Note that those results have not been symmetrized as we do in the SVD in order to show the robustness of the process.}
        \label{FigureSI_Trajectories}
    \end{center}
\end{figure}

We further validate our results with a numerical simulation performed with a finite elements software (Comsol Multiphysics). We mesh (the maximum element size is set to 3~mm) a 3-D virtual strip of dimensions $L\times w\times d=600\textrm{ mm}\times39\textrm{ mm}\times3\textrm{ mm}$ (similar to the experiment) and input the shear properties determined in the experiment. The point at a distance $w/6$ from the central axis is vibrated monochromatically in the direction ${x_1}+{x_2}$ in order to efficiently feed all the modes. Both the Dirichlet and Neumann boundary configurations are investigated (see Figure.~\ref{FigureSI_Simulations}).
The post-processing operations are the one used for the experiments (\textit{i.e.} decomposition over the symmetrical and anti-symmetrical parts, singular value decomposition and Fourier analysis). We find a very convincing matching over the whole 0-200~Hz frequency range.
Note how, just like for the experimental data, the Fourier analysis fails to extract the solutions corresponding to essentially imaginary wave numbers (see right part). This is particularly obvious in the area where the branches $S_1$ and $S_2$ (resp. $A_1$ and $A_2$) repel each other.

\FloatBarrier

\FloatBarrier

\FloatBarrier

\FloatBarrier

\FloatBarrier

\FloatBarrier

\FloatBarrier



\movie{Strip ($w=39$~mm) with free edges shaken at 110~Hz (same strip as in Fig.~2 of the main file). Left: raw acquisition. Center: Displament fields $u_1$ and $u_2$ after extraction by Digital Image Correlation (DIC). Right: Image with magnified motion ($\times35$) \text{via} the displacement data.}

\movie{Strip ($w=39$~mm) with free edges shaken at 110~Hz (same strip as in Fig.~2 of the main file). Displacement field maps and projection on the eigen modes.}

\movie{Strip ($w=39$~mm) with free edges shaken at 130~Hz (same strip as in Fig.~2 of the main file). The data were spatially filtered to isolate the backward ($S_2$) mode.}

\movie{Strip ($w=50.6$~mm) with fixed edges shaken at 110~Hz (same strip as in Fig.~3 and 4 of the main file). Left: raw acquisition. Center: Displament fields $u_1$ and $u_2$ after extraction by Digital Image Correlation (DIC). Right: Image with magnified motion ($\times35$) \text{via} the displacement data. The grey frame is symbolic.}

\movie{Strip ($w=50.6$~mm) with fixed edges shaken at 129~Hz (same strip as in Fig.~3 and 4 of the main file). Left: Field patterns at the Dirac point separated following the SVD procedure. Right: We select specific particles of the strips and follow their motion over a full wave cycle. The motion was magnified ($\times100$).}

\movie{Strip ($w=50.6$~mm) with fixed edges. The source is placed in the centre of the strip and shaken vertically at 136~Hz (corresponding to Fig.~4(a) in the main file). Here we report the field patterns of the symmetrical displacement (see fig. S7 for total displacement) as well as trajectories of specific particles over a full wave cycle. The motion was magnified ($\times150$).}

\movie{Strip ($w=50.6$~mm) with fixed edges. Two sources are facing each-other and rotated symmetrically at 136~Hz  (corresponding to Fig.~4(b) in the main file) for selective excitation of $S_1$. Here we report the field patterns of the symmetrical displacement as well as trajectories of specific particles over a full wave cycle. The motion was magnified ($\times150$).}

\movie{Strip ($w=50.6$~mm) with fixed edges. Two sources are facing each-other and shaken horizontally in an anti-symmetrical manner at 102~Hz (corresponding to Fig.~4(c) in the main file) for excitation of modes $A_1$ and $A_2$ in the lower region and $A_1^*$ and $A_2^*$ in the upper region of the strip. We report the field patterns of the anti-symmetrical displacement as well as trajectories of specific particles over a full wave cycle. The motion was magnified ($\times100$).}

\movie{Strip ($w=50.6$~mm) with fixed edges. Two sources are facing each-other and rotated in an anti-symmetrical manner at 102~Hz (corresponding to Fig.~4(d) in the main file) for selective excitation of modes $A_1$ in the lower region and $A_2^*$ in the upper region of the strip. We report the field patterns of the anti-symmetrical displacement as well as trajectories of specific particles over a full wave cycle. The motion was magnified ($\times30$).}



\bibliography{biblio}